\documentclass{article}
\usepackage{arxiv}

\usepackage[utf8]{inputenc} 
\usepackage[T1]{fontenc}    
\usepackage{hyperref}       
\usepackage{url}            
\usepackage{booktabs}       
\usepackage{amsfonts}       
\usepackage{nicefrac}       
\usepackage{microtype}      
\usepackage{lipsum}
\usepackage{fancyhdr}       
\usepackage{graphicx}       
\graphicspath{{media/}}     
\usepackage[numbers]{natbib}

\title{Speeding up charge exchange recombination spectroscopy analysis in support of NERSC/DIII-D realtime workflow
\thanks{\textit{\underline{Citation}}: 
\textbf{\href{https://arxiv.org/abs/2309.08687}{arXiv:2309.08687}}}
}
\author{
  Aarushi Jain\\
  University of Texas- Arlington\\
  Arlington, Texas\\
  \texttt{axj8943@mavs.uta.edu}
  \\\And
  Laurie Stephey\\
  Lawrence Berkeley National Lab\\
  Berkeley, California\\
  \texttt{lastephey@lbl.gov}\\
  0000-0003-3868-6178
  \\\And
  Erik Linsenmayer\\
  General Atomics\\
  San Diego, California\\
  \texttt{linsenmayere@fusion.gat.com}
  \\\And
  Colin Chrystal\\
  General Atomics\\
  San Diego, California\\
  \texttt{chrystal@fusion.gat.com}\\
  0000-0003-3049-8658
  \\\And
  Jonathan Dursi\\
  NVIDIA\\
  \texttt{jdursi@nvidia.com}
  \\\And
  Hannah Ross\\
  Lawrence Berkeley National Lab\\
  Berkeley, California\\
  \texttt{hross@lbl.gov}
}

\begin{document}
\maketitle

\begin{abstract}
We report optimization work made in support of the development of a realtime Superfacility workflow between DIII-D and NERSC. At DIII-D, the ion properties measured by charge exchange recombination (CER) spectroscopy are required inputs for a Superfacility realtime workflow that computes
the full plasma kinetic equilibrium. In this workflow, minutes matter since
the results must be ready during the brief 10-15 minute pause between plasma discharges. Prior to this work, a sample
CERFIT analysis took approximately 15 minutes. Because the problem consists of many calculations that can be done independently, we were able to 
restructure the CERFIT code to leverage this parallelism with Slurm job arrays. We reduced the runtime to approximately 51 seconds- a speedup of roughly 20x, saving valuable time for both the scientists interested in the CER results and also for the larger equilibrium reconstruction workflow. 
\end{abstract}


\section{Introduction}

DIII-D is currently the largest operating magnetic confinement fusion experiment in the United States. It operates over campaigns several months long, with 30-40 plasma discharges per experimental runday, with experimental discharges occurring approximately every 15 minutes. DIII-D has over 50 diagnostics \cite{noauthor_diii-d_2023-1}, many of which require moderate to substantial computational resources to convert from raw signals to usable physical measurement quantities. \cite{noauthor_diii-d_2023} 

The major goal of the DIII-D/NERSC Superfacility effort \cite{smith_vision_2023,bard2022lbnl} is to optimize an equilibrium reconstruction workflow at NERSC so that the walltime is fast enough to enable between-discharge analysis. This Superfacility work builds on previous efforts to couple magnetic confinement fusion experiments with realtime HPC resources \cite{kostuk_automatic_2018,kube_near_2022}. CAKE is Consistent Automatic Kinetic Equilibrium \cite{xing_cake_2021}- it is a module within the larger OMFIT framework \cite{meneghini_integrated_2015}. The ion properties measured by the CER (charge exchange recombination spectroscopy) diagnostic are required inputs for the CAKE workflow, so the reconstruction portion of CAKE cannot start running until CERFIT, the analysis suite for the CER diagnostic, completes. 

CAKE provides fully kinetic plasma equilibrium reconstructions, yielding a higher accuracy magnetic topology of the plasma. Although the plasma topology is important to both operators and researchers, it is computationally expensive to obtain and the current CAKE walltime is too long to be run between discharges. As a result, CAKE is often run after the experimental runday has completed; at this time any insights about plasma equilibrium are no longer actionable. 

\begin{figure}[thpb]
    \centering
    \framebox{\parbox{3in}{\includegraphics[width=3in, height=2.5in]{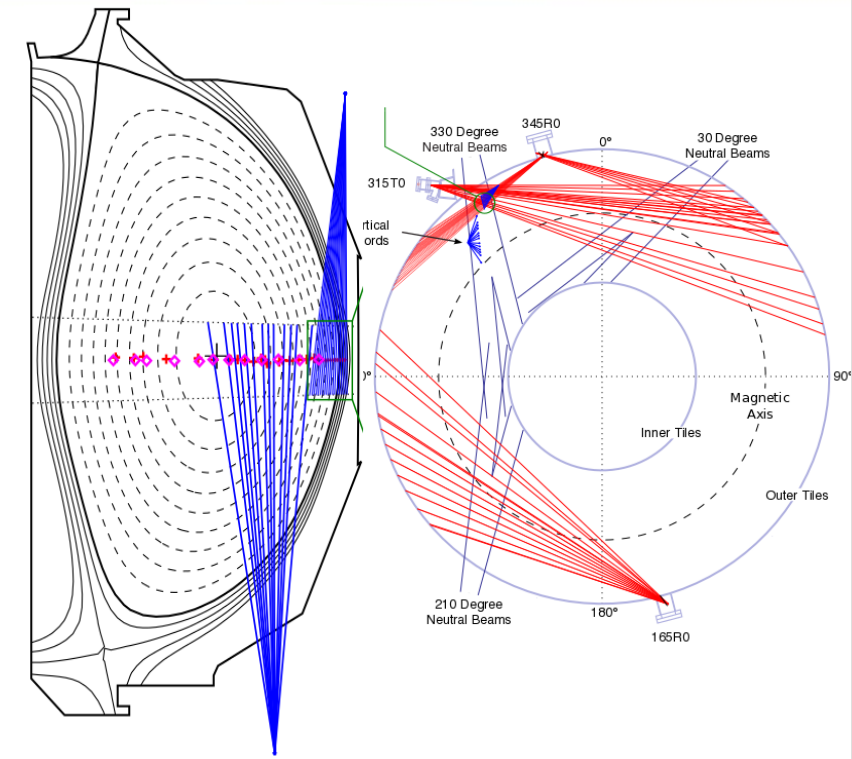}}}
    \caption{A diagram of the DIII-D Charge Exchange Recombination Spectroscopy diagnostic system (CER). There are approximately 76 total channels in the system; vertical viewing channels (chords) are shown in blue, and tangential viewing channels (chords) are shown in red. 64 chords are analyzed in standard CERFIT processing \cite{chrystal_improved_2016}.}
    \label{fig:chords}
\end{figure}

The DIII-D CER diagnostic system (see \cite{chrystal_improved_2016} and the references therein) is comprised of approximately 76 channels, each of which has a different viewing position, or chord, into the DIII-D vacuum vessel. A diagram of the CER system is shown in Figure \ref{fig:chords}. Visible light from these chords is analyzed by spectrometers, and particular spectral lines emitted by the plasma are used to determine ion velocity, temperature, and density via the measured Doppler shift, Doppler broadening, and radiance. These quantities are derived from fits to the spectra, and the code used to perform these fits is called CERFIT \cite{seraydarian_multichordal_1986,chrystal_improved_2016}. For standard analysis, CERFIT processes 64 chords.

The duration of a typical DIII-D discharge is between 4 and 10 seconds, and the CER system normally acquires data at 200 Hz. While any one spectral fit is not computationally difficult, complete analysis for a typical discharge requires more than 10,000 fits, and executing the highest quality automatic version of these fits in series previously required approximately 15 minutes running on a node of the DIII-D Omega cluster (node details are discussed in detail in Section 2). Since information from a previous discharge is needed as quickly as possible to inform the setup of the next discharge, automatic CER fits done between discharges are less complex (and correspondingly less accurate) and take approximately 4 minutes to complete. For this work, the goal was to speed up the highest quality fitting (which is required for CAKE) by at least a factor of 10 so that those results could be used between discharges and also be inputs to further computations that also aim to complete significantly before the next discharge. 

\section{Experimental setup}

We performed this study using the local DIII-D cluster, Omega. Owing to its nature as a data-analysis code, CERFIT is substantially less portable than a simulation code. CERFIT has complex dependencies that are not easily satisfied on NERSC systems. One of these dependencies is a custom DIII-D data access library \texttt{PTDATA} \cite{mcharg_jr_diii-d_2002}, which to our knowledge has not yet been installed on a system outside of DIII-D. In addition to the \texttt{PTDATA} library, the DIII-D raw data itself much also be accessible as an input to CERFIT, including the timing and profile of neutral beams, which requires access to the DIII-D data system. CERFIT has several sub-modules which are tracked via a set of environment variables, which in turn can be set by custom system modules. The CERFIT test suite requires known-good datasets in an expected directory structure. All of these factors make it a challenge to move CERFIT to another system; however, we would like to study the feasibility of running this workflow at NERSC to leverage the additional computational resources available on Perlmutter as future work \cite{NERSC_perlmutter}. 

The Omega cluster is a heterogeneous Linux cluster situated at DIII-D. It contains two login nodes with 2x Intel Xeon Gold 6252 CPUs and 34 total compute nodes including: 2 nodes of 2x Intel Xeon Gold 6252 CPUs, each with 1 NVIDIA V100, 1 node of 2x Intel Xeon Gold 6252, 2 nodes of 2x Intel Xeon Platinum 8260, 16 nodes of 2x AMD EPYC 7513, 12 nodes of 2X AMD EPYC 7502, and 1 node of 2x AMD EPYC 7343. Job submission is managed by the Slurm scheduler. For queue permission reasons, we have targeted the two Intel Xeon Gold 6252 CPUs nodes during our study, as the CERFIT code currently must be compiled and run on Intel nodes due to an apparent outstanding nvfortran compiler issue on AMD nodes. Work is ongoing to address this nvfortran issue. To work around this, the CER team is actively switching to the gfortran compiler. Early results have shown that there are no issues using the AMD hardware with gfortran.

\section{Determining Optimization Strategy}

FITCER, the outer wrapper for CERFIT, performs complete CERFIT analysis of all CER chord data from a discharge. The CERFIT analysis code is comprised both of Fortran and C code- the main analysis algorithms are written in Fortran, and the main data access components are written in C. CERFIT performs spectral fitting by adding Gaussian functions to produce a synthetic spectra (\textit{Sum of Gauss}), and then using the Levenberg-Marquardt minimization algorithm, CERFIT iteratively fine-tunes the model to minimize the \(\chi^2\) value of the fit, ultimately obtaining the best fit ion properties based on the fit line's location, width, and amplitude (\textit{Chi-squared minimization}). There are several additional data acquisition and pre-processing steps.  

We began this optimization study with application profiling. Understanding the structure of the CERFIT application and where the time was spent was crucial to identifying a strategy to achieve speedup. We used the NVIDIA Nsight system profiler \texttt{nsys} for our CPU code since that was available on the system. We profiled the application running on a single process/single CPU since that was how CERFIT was typically run. Our main interest was how much speedup can be practically achieved for the team given the conditions under which they typically run.

\begin{figure}[thpb]
    \centering
    \framebox{\parbox{3in}{\includegraphics[width=3in, height=2.5in]{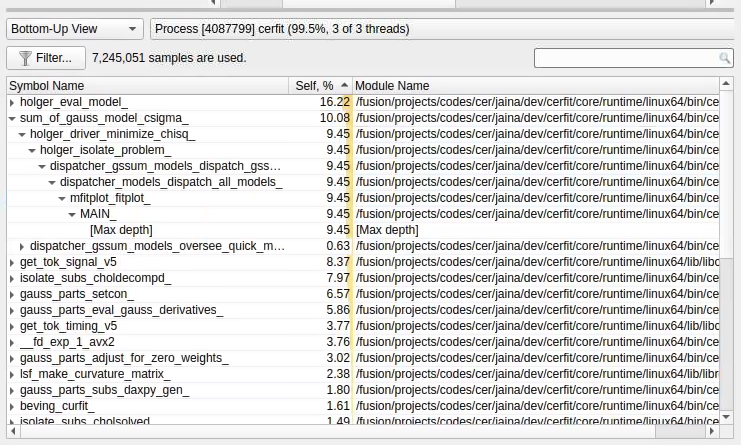}}}
    \caption{NVIDIA \texttt{nsys} profiling results illustrating the runtime distribution of the CERFIT code in ``Bottom-up view''. Note that \texttt{nsys} is displaying the symbol names rather than the function names. The \textit{Chi-squared minimization}, symbol \texttt{holger\_eval\_model}, occupies approximately 16\% of the total runtime, while the \textit{Sum of Gauss} operations, symbol \texttt{sum\_of\_gauss\_model\_csigma}, used for preparing synthetic spectra, accounts for around 10\%. The absence of a single hotspot suggests that GPU porting of these algorithms may not yield significant overall speedup without a major redesign of the CERFIT code.}
    \label{fig:nsys}
\end{figure}

\begin{sloppypar}
Sample NVIDIA \texttt{nsys} profiling data are shown in Figure \ref{fig:nsys}. Displaying the nsys profiling data in ``Bottom-up'' view \cite{nsys_modes} provided useful information for our CPU-only application. It provided information about the structure of the application and the major hotspots in the code. The \textit{Chi-squared minimization}, symbol \texttt{holger\_eval\_model}, took about 16 percent of total runtime, and the \textit{Sum of Gauss} function, symbol \texttt{sum\_of\_gauss\_model\_csigma}, that constructed synthetic spectra for fitting took about 10 percent of total runtime. Many individual functions performing adjacent tasks accounted for the remainder of the application runtime, which we believe is a result of application branching. 
\end{sloppypar}

These profiling data indicated that no single function was doing a lot of the heavy lifting in CERFIT. In a best-case scenario in which we ported both \textit{Chi-squared minimization} and \textit{Sum of Gauss} to GPU and achieved speedup, we would still be speeding up only about a quarter of the application runtime. Although our initial goal in this study was to try to adapt CERFIT to leverage GPUs, the profiling data made it clear that CERFIT in its present state was not a good candidate for GPU optimization.

Recognizing the limited potential from GPU porting, we redirected our focus towards obtaining CPU speedup through adding parallelism. Both from discussions with the CER team and from our profiling data, we were aware that CERFIT was performing fully independent calculations for each CER chord and could be reconfigured to run in an embarrassingly parallel manner. The independent nature of the calculations made the algorithm an excellent candidate for parallelization, so we decided to pursue this approach. 

\section{Implementing chord-level CPU parallelism}

\begin{figure}[thpb]
    \centering
    \framebox{\parbox{3in}{\includegraphics[width=3in, height=2.5in]{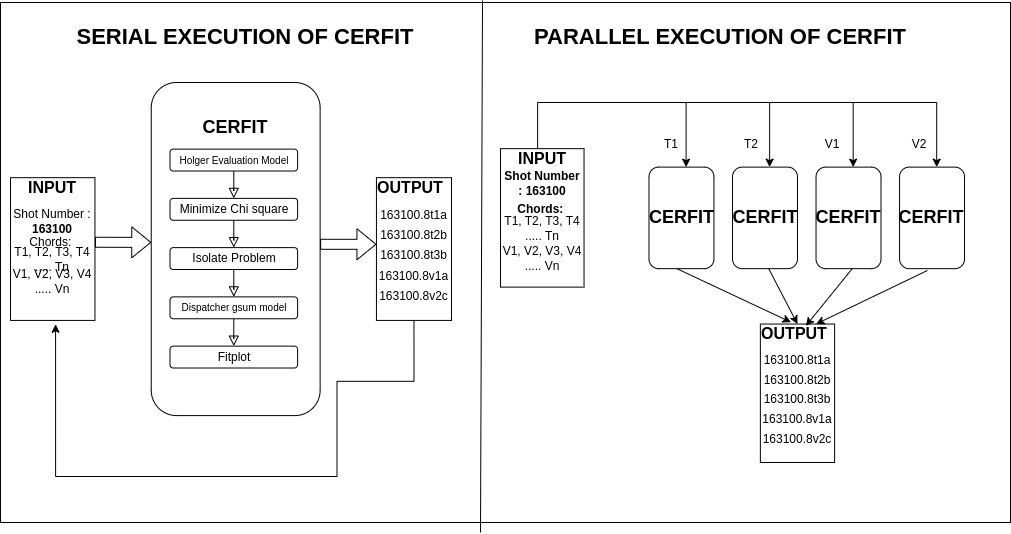}}}
    \caption{Illustration of serial and parallel execution of CERFIT. In the serial execution, a single input file contains data for all chords, and CERFIT processes each chord sequentially. In contrast, the parallel execution divides the input data into separate files, one for each chord. Multiple instances of CERFIT run in parallel in this embarrassingly parallel model.}
    \label{fig:cerfit_diagram}
\end{figure}

Given the complex nature of the CERFIT code (see Section 3), which involves multiple branching points, our goal was to validate our understanding of chord-level parallelism. We first needed to locate the right place in the code where the chord-splitting takes place. Once located, as a test, we used the Linux utility \texttt{xargs} to divide the CERFIT input file in two pieces and launch two independent CERFIT jobs, each running on a subset of the input file. We observed that CERFIT ran as expected in this mode of operation and determined it was safe to move forward with this strategy depicted in Figure \ref{fig:cerfit_diagram}. 

To obtain results quickly, we began by writing a prototype Python script to break the CERFIT input file into 64 individual input files- one for each viewing chord. This prototype script enabled us to explore large-scale parallelization via Slurm job arrays. Slurm job arrays \cite{slurm_array} are a feature in the Slurm workload manager that enables many similar jobs to be launched from a single batch script. We decided to start with Slurm job arrays instead of a solution like MPI largely due to the lower barrier to entry and faster path to implementation. Incorporating MPI into CERFIT, a complex C and Fortran code, would likely be time-consuming and can be considered for future work. One benefit of relying on Slurm to provide parallelism is that the distribution of work is extremely flexible and the size of the job can be quickly adjusted (i.e. using one node instead of two). This is one major advantage compared to a similar MPI implementation. 

\begin{sloppypar}
To enable FITCER to launch its own Slurm jobs, we wrote a Slurm job array script designed to launch 64 nearly identical CERFIT jobs, one for each CER chord. The variable \texttt{SLURM\_ARRAY\_TASK\_ID} was used to access the corresponding CERFIT chord input file. The script launched one job per physical CPU core with 24 total jobs per node. Additionally, it was necessary to allocate the memory per CPU correctly so that the resources requested for each job would not exceed the total resources on the node and cause the jobs to block. With this configuration, we observed that our jobs ran in parallel as expected. Our Slurm job array batch script is shown below:
\end{sloppypar}

\begin{verbatim}
#!/bin/bash
#SBATCH -p gpus
#SBATCH --array=1-64
#SBATCH --cpus-per-task=1
#SBATCH -n 1
#SBATCH --ntasks-per-node=24
#SBATCH --mem-per-cpu=1G

srun time cerfit < chord $SLURM_ARRAY_TASK_ID.in \
>& fit_$SLURM_ARRAY_TASK_ID.out
\end{verbatim}

Once we confirmed that this new chord-parallel structure was suitable, our objective was to implement our changes back into the production version of FITCER, with the goal of making the parallel version of the code look and feel like the original version. The key feature was to re-implement the chord-splitting function that had been tested via a Python script into FITCER in Fortran. A specialized Fortran subroutine was developed to effectively partition individual chords from the primary input file called \texttt{fit\_<chord\_number>.in}. One challenge in translating from a Python prototype to a Fortran subroutine is that handling multi-file I/O requires more care and certainly more lines of code. Another challenge was handling the splitting in a robust manner. During the process, we observed that the structure of the CERFIT input file differs between production runs and running within the test suite. As a result, this meant that we needed to split the file on a certain word rather than after a certain number of lines. This resulting subroutine seamlessly integrates within the existing \texttt{fitcer\_input\_file.f90} code.  

\section{CER Regression Testing}

Verifying that our modifications to CERFIT did not meaningfully alter the outputs of the code was essential. We used the established CERTEST test script developed at DIII-D to check for correctness of our CERFIT outputs. 

CERTEST can be run in two modes- in the first mode, it generates a set of known-good reference files stored in a bespoke directory. In test mode, CERTEST runs CERFIT and generates standard output files, which are then compared to the reference files. We generated the reference files using the original serial code, and we generated the test files using our parallel version of CERFIT. CERTEST needed some modifications to be able to handle the paradigm in which output files are written from Slurm jobs. Once we ran the updated CERTEST, the CER team determined that the differences that were present between the two versions were negligible. 

\section{Results}

In this section we discuss the measurements to assess the overall improvements in runtime from our implementation of chord-level parallelism via Slurm job arrays.

\begin{figure}[thpb]
    \centering
    \framebox{\parbox{3in}{\includegraphics[width=3in, height=2.5in]{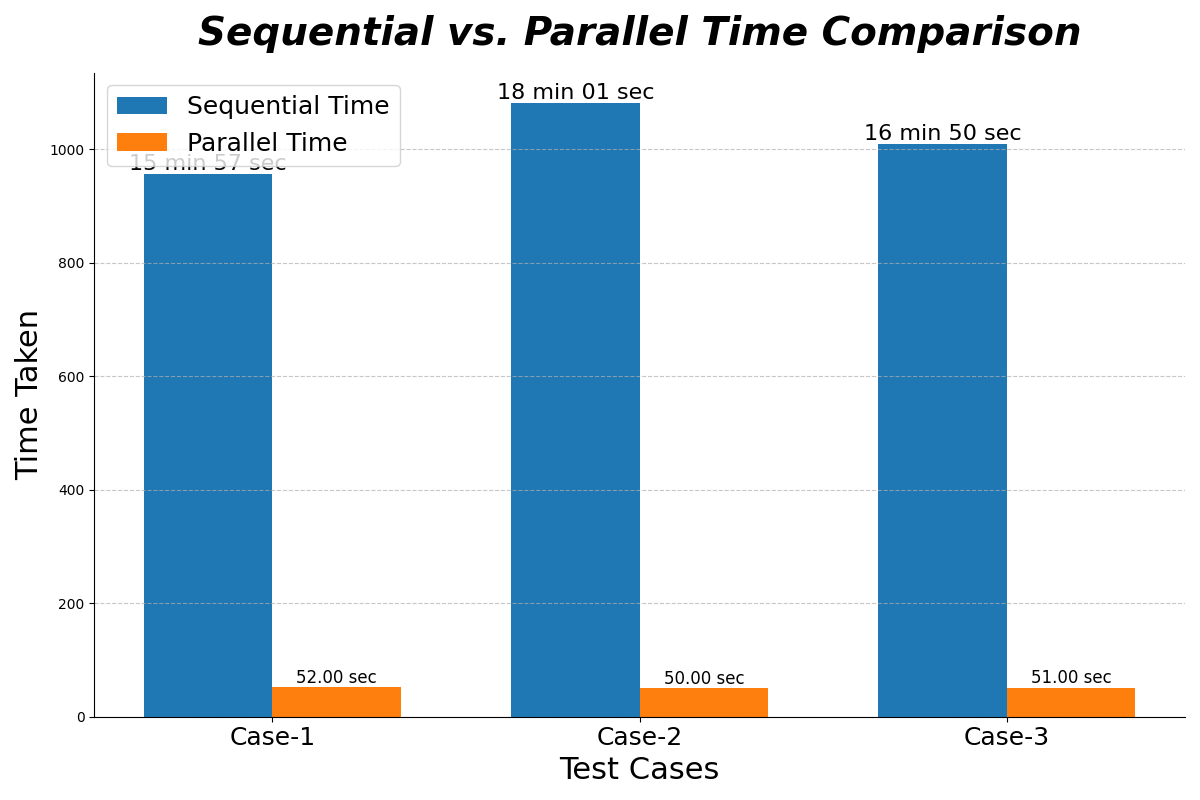}}}
    \caption{Benchmarking results of sequential vs. parallel CERFIT Execution. This plot presents a comparison of execution times for 3 trials of each case. Sequential execution of FITCER, indicated by the blue bars, completes in approximately 15 minutes. In contrast, parallel execution, indicated by the orange bars, completes in approximately 50 seconds. On average, the parallel version yields a 20x speedup relative to the sequential version.
}
    \label{fig:benchmarking}
\end{figure}

We used the following procedure to perform our benchmarking on the Omega cluster and achieve our speedup results shown in Figure \ref{fig:benchmarking}:
\begin{enumerate}
\item We ran CERFIT via FITCER using a standard sample discharge 163100 for analysis. This run was performed on an Omega Intel Xeon Gold 6252 CPU login node since this is the standard procedure for the CER team. We ran 3 trials of the original serial implementation, shown in blue, using the Linux \texttt{time} command to obtain the measurement. The mean runtime of these 3 trials was 1010 seconds. 
\item We ran our parallelized version of FITCER using the same standard sample discharge. For this benchmark we ran on two Omega Intel Xeon Gold 6252 compute nodes, each with 24 physical cores, for a total of 48 physical cores. 
We ran 3 trials of the parallel implementation, shown in orange, using a special bash wrapper script to obtain the full FITCER runtime. It was not adequate to time FITCER using Linux \texttt{time} since FITCER currently finishes executing as soon as the job array is submitted via SBATCH. In the wrapper script, we included a while loop to query Slurm every 2 seconds to determine if the job array jobs were still running. When the query returned no results, the timer ended and the runtime duration was calculated. Future work could include adding such timing capabilities into FITCER itself. The mean runtime of these 3 trials was 51 seconds. Dividing the original 1016 seconds by 51 seconds, we arrive at a speedup of 19.9, or approximately 20x.
\end{enumerate}

\begin{figure}[thpb]
    \centering
    \framebox{\parbox{3in}{\includegraphics[width=3in, height=2.5in]{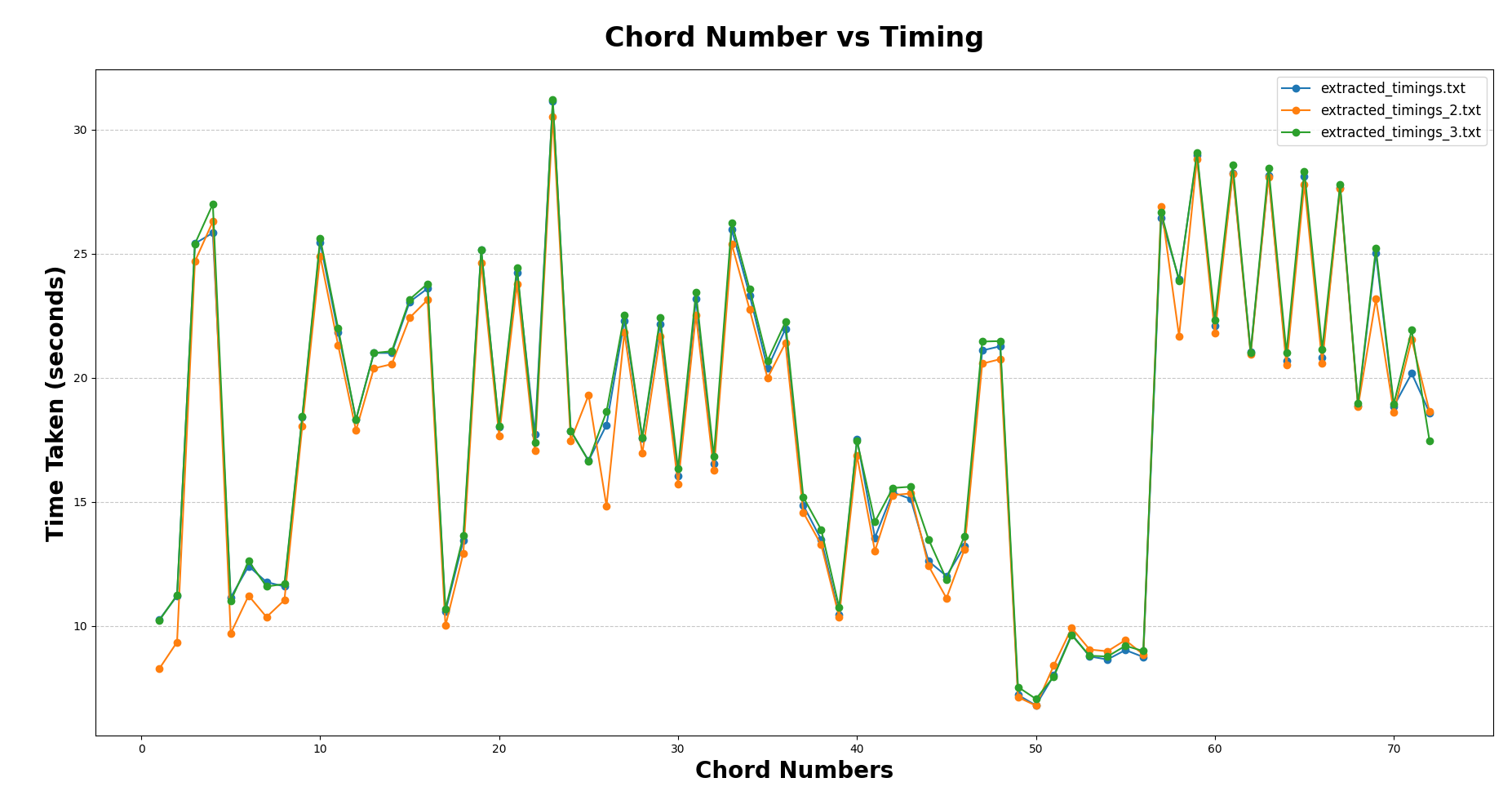}}}
    \caption{Examination of the individual chord runtimes in CERFIT. We performed 3 trials of this test. The plot shows chord number vs execution time, with each trial shown in a different color. We observed that the differences in processing time between chords is robust.
}
    \label{fig:each_chord}
\end{figure}

We performed a separate study to examine the individual chord processing times. The execution times for each chord are shown in Figure \ref{fig:each_chord}. We ran 3 trials of this CERFIT per-chord test and found that some chords are consistently processed more quickly than others. Given this behavior, using a flexible framework like Slurm job arrays affords some load-balancing by allowing any cores that many have finished early the opportunity to process another channel. 

It is additionally useful to discuss the speedup in terms of a per-core analysis. The initial benchmark (roughly 1016 seconds) was run in serial on a single core. The final benchmark (roughly 51 seconds) was run on 48 physical cores. With perfect strong scaling one might expect a 48x speedup (roughly 19 seconds)- what accounts for this difference? First, there are 64 total chords being processed, so 16 CPUs processed more than one chord, which reduces the potential strong-scaling speedup. Next, Figure \ref{fig:each_chord} demonstrates that the time to process each chord varies from a few seconds to nearly 30 seconds. In an ideal situation, a CPU that finishes early with an ``fast'' chord would be available to process another chord. Using Slurm Job Arrays does inherently provide some load balancing in this regard, although it is not optimized. Finally, there is some overhead of submitting each CERFIT task as a Slurm job. We should note that the target CPU compute nodes were unoccupied, so we expect that there is some overhead to submit and start the job, but there was no queue wait-time (except from the application itself). If the test could have been performed on 3 nodes with more than 64 available CPU cores, we might expect to achieve closer to the ideal strong-scaling speedup, although the speedup will be determined by the ``long pole in the tent'', which in this case is the chord that takes approximately 30 seconds to be processed. Since there is so much variation in the per-chord processing time, achieving the ideal strong-scaling speedup of 19s is not possible with this application. 

\section{Future Work}

The work we have reported here is an initial effort towards speeding up the walltime of CERFIT analysis at DIII-D. However, there are many additional avenues for this work. 

First and foremost, the objective of this work is to develop a production quality parallel version of CERFIT that the CER and CAKE teams can use in routine analysis. Work is ongoing to finalize changes, including switching from nvfortran to gfortran and adjusting the Slurm job array configuration to fit Omega queue policies, to enable CERFIT to run routinely between discharges on the AMD compute nodes in the Omega cluster. 

One potential next step would be to study replacing the Slurm job array based parallelism with MPI. This would provide a scheduler-independent framework for parallelism. It would also be instructive to evaluate the differences in overhead between starting an independent Slurm job for each chord as compared to MPI process startup within a single job.  

The whole of this work was performed on the Omega cluster at DIII-D. However, another major area of study would be to try running this analysis on NERSC's Perlmutter. This would require some additional libraries (like \texttt{PTDATA}, a DIII-D internal data system library) to be locally installed at NERSC. It may also require some examination of the efficiency of external data transfer via \texttt{PTDATA}. The goal would be to determine if the realtime resources NERSC could offer would benefit the overall workflow and overcome the additional overheads of raw data transfer (estimated to be a relatively modest 1 GB per discharge). Running CERFIT locally at NERSC would mean that the ion physics outputs from CERFIT would be located at NERSC and could be used directly in CAKE and other workflows, which could be beneficial. Questions about how to locally store these data for efficient access and for what duration would need to be explored.

\section{Summary}

To summarize, this work describes our efforts to achieve 20x speedup for the high-quality CERFIT CER diagnostic analysis code used at DIII-D which yields plasma ion properties. This speedup is expected to benefit DIII-D scientists and operators working with CER results in the control room since it will provide them this high-quality information substantially faster. We believe this CERFIT speedup will additionally translate into speedup for the CAKE Superfacility workflow project that connects DIII-D to NERSC with the goal of enabling routine between-discharge plasma equilibrium reconstruction.

\section*{Acknowledgments}

This research used resources of the National Energy Research Scientific Computing Center (NERSC), a U.S. Department of Energy Office of Science User Facility located at Lawrence Berkeley National Laboratory, operated under Contract No. DE-AC02-05CH11231 using NERSC award ASCR-ERCAP0019913.

This material is based upon work supported by the U.S. Department of Energy, Office of Science, Office of Fusion Energy Sciences, using the DIII-D National Fusion Facility, a DOE Office of Science user facility, under Award(s) DE-FC02-04ER54698.

Disclaimer: This report was prepared as an account of work sponsored by an agency of the United States Government. Neither the United States Government nor any agency thereof, nor any of their employees, makes any warranty, express or implied, or assumes any legal liability or responsibility for the accuracy, completeness, or usefulness of any information, apparatus, product, or process disclosed, or represents that its use would not infringe privately owned rights. Reference herein to any specific commercial product, process, or service by trade name, trademark, manufacturer, or otherwise does not necessarily constitute or imply its endorsement, recommendation, or favoring by the United States Government or any agency thereof. The views and opinions of authors expressed herein do not necessarily state or reflect those of the United States Government or any agency thereof.

This work was completed in part at the July 2023 NERSC Open Hackathon, part of the Open Hackathons program. The authors would like to acknowledge OpenACC-Standard.org for their support. 

The authors would like to thank David Schissel and 3 anonymous reviewers from the SC23 XLOOP workshop for their suggestions to improve this paper. 

Finally, the authors would like to thank the CAKE Superfacility team for useful discussions that contributed to this work.

\bibliographystyle{hunsrtnat}  
\bibliography{cerfit}

\end{document}